\documentclass[12pt]{article}
\usepackage{a4wide}
\usepackage{amsmath}
\usepackage{amssymb}
\usepackage{graphicx}
\usepackage{latexsym,epsfig,cite}

\newcommand{\mg}{m_{\tilde{G}}}
\newcommand{\tg}{\tau_{\tilde{G}}}

\def\beq{\begin{equation}}
\def\eeq{\end{equation}}
\def\bea{\begin{eqnarray}}
\def\eea{\end{eqnarray}}

\def \mm {m_{\tilde G}}

\def\lsim{\mathrel{\rlap{\raise 2.5pt \hbox{$<$}}\lower 2.5pt\hbox{$\sim$}}}
\def\gsim{\mathrel{\rlap{\raise 2.5pt \hbox{$>$}}\lower 2.5pt\hbox{$\sim$}}}

\allowdisplaybreaks 

\usepackage[dvips]{color}
\definecolor{Black}{named}{Black}
\definecolor{Red}{named}{Red}

\begin{document}

\thispagestyle{empty}
\begin{flushright}

\end{flushright}
\vspace*{5mm}
\begin{center}
{\large {\bf Photon, Neutrino and Charged Particle
Spectra from R-violating Gravitino Decays}}\\
\vspace*{1cm}
{\bf N.-E.\ Bomark$^1$, S.\ Lola$^2$, P.\ Osland$^1$} and
{\bf A.R.\ Raklev$^{3}$} \\
\vspace{0.3cm}
$^1$ Department of Physics and Technology, University of Bergen,
N-5020 Bergen, Norway \\
$^2$ Department of Physics, University of Patras, GR-26500 Patras, Greece \\
$^3$ Oskar Klein Centre for Cosmoparticle Physics, Department of Physics,
Stockholm University, SE-10691 Stockholm, Sweden

\end{center}

\begin{abstract}
We study photonic, neutrino and charged particle signatures from slow
decays of gravitino dark matter in supersymmetric theories where
R-parity is explicitly broken by trilinear operators. Photons and
(anti-)fermions from loop and tree-level processes give rise to
spectra with distinct features, which, if observed, can give crucial
input on the possible mass of the gravitino and the magnitude and
flavour structure of R-violating operators. Within this framework, we
make detailed comparisons of the theoretical predictions
to the recent experimental data from PAMELA,
ATIC and Fermi LAT.

\end{abstract}



\setcounter{page}{1}

\section{Introduction}

In recent years, there has been a revived interest in the possibility of gravitino dark matter within the framework of R-violating supersymmetry~\cite{BM,LOR}. There are several reasons for this. In the much-studied CMSSM \cite{Chamseddine:jx}, scenarios with neutralino dark matter have become rather constrained, requiring a considerable amount of fine-tuning. 
The additional fact that R-parity is imposed essentially by hand, and alternative schemes are equally viable from the theoretical point of view, has motivated the search for other dark matter candidates. In the case that R-parity is violated, neutralinos and other sparticles decay too fast to be dark matter, even for small couplings, however, this is not the case for gravitinos, which raise very interesting possibilities. Indeed, if gravitinos decay slowly enough for their lifetime to exceed the age of the universe~\cite{TY,CM}, they can be considered essentially stable and a good dark matter candidate. While cosmologically stable, gravitinos may still have decays of interest to astrophysics measurements of cosmic rays in various particle species.

Trilinear R-violating couplings have the form
\begin{equation}
\lambda L_{i}L_{j}{\bar{E}}_{k}
+\lambda ^{\prime }L_{i}Q_{j}{\bar{D}_{k}}
+\lambda ^{\prime \prime }{\bar{U}_{i}}{\bar{D}_{j}}{\bar{D}_{k}},
\label{Rviol}
\end{equation}
where $L(Q)$ are the left-handed lepton (quark) doublet superfields,
and ${\bar{E}}$ (${\bar{D}},{\bar{U}}$) are the corresponding
left-handed singlet fields.  Due to SU(2) and SU(3) invariance, we
have 9 $L_iL_j{\bar{E}_k}$ operators $(i\neq j)$, 27
$L_iQ_j{\bar{D}_k}$, and 9 ${\bar{U}_i}{\bar{D}_j}{\bar{D}_k}$
operators $(j \neq k)$. These couplings are known to give rise to very
interesting collider phenomenology, with the missing energy signature
of the much studied Minimal Supersymmetric Standard Model (MSSM) being
in part substituted by multi-lepton and/or multi-jet events~\cite{barb}.

The same R-violating operators that lead to the decay of sparticles
produced in collider experiments, will also cause gravitino dark
matter to decay. This takes place via two-body radiative loop decays
to neutrino and photon~\cite{LOR}, and via tree-level decays to three
fermions~\cite{CM}. The decay rates have been presented in detail in
the original references, where it was found that, for light gravitino
masses and appropriate fermions in the loop, radiative decays may
dominate, while for heavier gravitinos, three-body decays take
over. The behavior of the total decay rate is primarily controlled by
the gravitino mass dependence of the partial decay widths. These scale as
\begin{align}
\Gamma_{\tilde G} & \propto  m_{\tilde G}^7 \quad\text{(three-body decays),}
\label{dec1}\\
\Gamma_{\tilde G} &\propto  m_{\tilde G}  \quad\text{(radiative decays),}
\label{dec2}
\end{align}
for three-body decays away from the kinematic threshold\footnote{This
  power of $m_{\tilde G}^7$, which plays an important role in our
  considerations, comes about as follows: there is a factor of
  $m_{\tilde G}^5$ from phase space, such as in muon decays, in
  addition, there is a factor of $m_{\tilde G}$ in the matrix element,
  since the gravitational coupling is proportional to the
  four-momentum.}, and radiative decays with gravitino masses well below
those of the next-to-lightest supersymmetric particle (NLSP) and
sufficiently above the kinematic threshold of its decay
products. Moreover, the results are very dependent on the flavour
structure of R-violating operators, which is discussed extensively
in~\cite{BLOR}.

To constitute a realistic dark matter candidate, the minimal
requirement is that the gravitino lifetime should exceed the age of
the universe, which may naturally occur due to the Planck scale
suppression in the gravitino vertex, the smallness of the
R-parity-violating coupling, and the additional loop/phase-space
suppression in the radiative/tree-level diagrams. In addition,
cosmic rays from slow gravitino decays have to be consistent with
observations, {\it e.g.} the photon flux measured in the EGRET
experiment~\cite{EGRET}. The extra-galactic photon flux has been shown
to set severe bounds on gravitino decays and thus on the allowed
combinations of gravitino masses and R-violating couplings
\cite{BM,LOR,BLOR,IbTran}. Recently, new results on the flux of
charged particles have been published by the ATIC~\cite{ATIC},
PAMELA~\cite{PAMELAep,PAMELApbar}, and Fermi LAT
Collaborations~\cite{Abdo:2009zk}. These show possible excesses that,
modulo unknown astrophysical sources, could signal New Physics.

One could also consider the possibility of tree level decays to a neutrino and photon through neutrino--neutralino mixing. This decay width is directly proportional to the R-parity violating operators' contribution to the neutrino masses. For a given choice of basis \cite{barb}, this in turn depends on the alignment between the vevs of the neutral scalar components of the superfields $(H_d,L_i)$ and the coefficients of the bilinear terms $(\mu,\mu_i)$, where $\mu_iH_uL_i$ are potentially non-zero bilinear R-violating operators.

The expected charged particle spectra of mixing induced decays have recently been studied in detail in \cite{Chpart,Buchmuller:2009xv,Ishiwata:2008cu,Ishiwata:2008cv}, and the neutrino spectra in \cite{Nspectra}. The alignment angle is very restricted in order to comply with neutrino mass bounds, and a completely general phenomenological model allows these decays to be sub-dominant to the decays discussed above. Thus, our present Letter complements these analyses for the case of decays dominated by trilinear R-violating operators\footnote{For a model of R-parity violation with gravitino decay channels similar to ours, see~\cite{Chen:2009ew}.}. As we will demonstrate, under certain assumptions on the background electron spectrum, it is possible to simultaneously fit the PAMELA data on the positron and antiproton fractions and the Fermi LAT data on the electron-plus-positron spectrum. Such a fit requires high gravitino masses, and the resulting gamma ray spectrum predicted from electron bremsstrahlung should be testable with the upcoming Fermi LAT data. For the neutrinos we find that their flux will be very difficult to detect in both present and near-future experiments.

This Letter is structured as follows: In Section~\ref{sec:charged}, we
study charged fermion and anti-fermion spectra from gravitino decays
using a standard galactic propagation model. We attempt a fit to the
PAMELA and Fermi LAT data, discussing the limitations of the
background model, and we set limits on the R-parity violating
couplings as a function of the gravitino mass. In
Section~\ref{sec:photons}, after discussing our procedure and the
handling of experimental uncertainties, we proceed to a study of
photonic spectra; we discuss the information that can be obtained from
characteristic peaks directly linked to the magnitude and flavour of
the R-violating couplings, the mass of the gravitino and the SUSY
spectrum. Furthermore, we make predictions corresponding to our best
fits to the PAMELA and Fermi LAT data, and show that these can be
tested in the near future, either supporting our considerations, or
setting even more severe bounds on R-violating couplings. In
Section~\ref{sec:neutrinos} we discuss the potential for observation
of the neutrino spectra from gravitino decays. Finally in
Section~\ref{sec:conclusions} we summarise our results and look at
future prospects.

\section{Charged particles}
\label{sec:charged}
\setcounter{equation}{0}

Gravitinos decaying through the tree level diagrams that are induced
by trilinear R-violating operators, will produce charged particles,
which, for appropriate values of the gravitino mass and the
R-violating couplings may help to explain recent experimental data on
cosmic rays. Of particular interest in this respect is the
measurement of the cosmic ray positron fraction by
PAMELA~\cite{PAMELAep} and the Fermi LAT results~\cite{Abdo:2009zk} on
the cosmic ray electron spectrum that both show excesses, but in somewhat
different energy ranges. In addition, the non-observation of any
anomaly in the antiproton data as reported by PAMELA\cite{PAMELApbar}
is important in constraining the possible parameter space.

\subsection{Model}

We calculate the electron, positron and antiproton spectra in the
gravitino rest frame for a particular gravitino mass and R-violating
coupling. We study the decay of gravitinos using
{\sc PYTHIA 6.4}~\cite{PYTHIA}, and the branching ratios and lifetime given by
the formulae in~\cite{CM,BLOR}. After being produced, charged
particles will propagate through the galaxy. Although we have
substantial information about the various processes that give important
contributions, this propagation still contains large uncertainties. For a
recent review of cosmic ray propagation, see~\cite{CRRev}.

The most important propagation effects come from diffusion processes
caused by galactic magnetic fields and scattering off
magnetohydrodynamic (MHD) waves; this spatial diffusion causes the
received spectra to be essentially isotropic. The scattering off MHD
waves also causes diffusion in momentum space known as diffusive
re-acceleration. It is also possible that an important role is played
by convection due to galactic winds.  At present, it seems that the
most appropriate model for galactic propagation is a diffusion model,
possibly extended with some convection~\cite{CRRev}.  On top of this,
charged particles are losing energy due to synchrotron radiation,
bremsstrahlung, ionization and inverse Compton scattering.

To take all these effects into account, we use the {\sc GALPROP}
code~\cite{GALPROP} with a conventional diffusion model, similar to
`model 0' in~\cite{Grasso:2009ma}. The main parameters adopted for GALPROP
are given in Table~\ref{table:GALPROP params}.  The dark matter
component of the various fluxes incident on the earth has been
calculated assuming a Navarro-Frenk-White (NFW) \cite{NFW} dark matter
density profile for our galaxy:
\begin{equation}\label{Eq:NFW}
  \rho_\text{Halo}(r)=\frac{\rho_0}{(r/r_c)(1+r/r_c)^2},
\end{equation}
where $r$ denotes the distance to the center of the galaxy,
$r_c = 20$~kpc and $\rho_0 = 0.33$~GeV~cm$^{-3}$.

\begin{table}[h]
\begin{center}
{
\begin{tabular}{|c|c|c|c|c|c|}
\hline
$D_0~[\text{cm}^2\text{ s}^{-1}]$ & $\delta$ &  $z_h~[\text{kpc}] $ & $\gamma_0$ & $N_{e^-} ~[\text{cm}^{-2}\text{ s}^{-1}\text{ sr}^{-1}\text{ GeV}^{-1}]$  &  $\gamma_0^P$\\\hline
$5.75\times 10^{28} $ & 0.34 & 4 & 2.50 & $4.0\times 10^{-7}$ & 2.36 \\
\hline
\end{tabular}
}
\caption{\label{table:GALPROP params} \it GALPROP parameters. With the
  diffusion coefficient $D$ assumed to have a power-law dependence on
  the energy, with index $\delta$, $D_0$ denotes its value at $4$ GeV;
  $z_h$ is the half-height of the halo in which the cosmic rays are
  assumed to propagate. Finally, $\gamma_0$ and $N_{e^-}$ give the
  spectral index and normalization (at $34.5$ GeV) of the electron
  injection spectrum, while $\gamma_0^P$ is the spectral index of the
  nuclei injection spectra.  }
\end{center}
\end{table}

At low energy, below 5--10~GeV, the resulting background spectra for
both positrons and antiprotons in this model deviate somewhat from the
data. This is believed to be due to solar modulation
effects~\cite{PAMELAep}, and not any potential New Physics. For the
positron fraction, therefore, only points above 10 GeV are used for
our fits\footnote{The question of exclusion or inclusion of individual
  points in the low energy region is important because the error bars
  here are smaller than in the higher energy region.}.

\subsection{Positrons and electrons}

Cosmic ray electrons and positrons are usually divided into primary cosmic rays originating in cosmic accelerators such as {\it e.g}. supernova remnants (SNR), and secondary cosmic rays stemming from cosmic rays (mostly protons)
interacting with the interstellar medium~\cite{Moskalenko:2004fe}. Positrons are,
unlike electrons, believed to originate mostly from secondary cosmic rays~\cite{CRRev}.
At the energies of interest, the secondary cosmic ray particle spectrum can be assumed to be charge symmetric, with
the secondary positrons and electrons having the same spectrum.
The primary electrons, together with secondary electrons and positrons, are
expected to give an exponentially falling electron-plus-positron spectrum,
as well as positron fraction, at high energy~\cite{Grasso:2009ma}.

Recently, several deviations from this picture have been observed in cosmic ray
electron and positron data. The PAMELA Collaboration~\cite{PAMELAep}
has observed a steep rise in the positron fraction above 10~GeV, with the
data continuing up to 100~GeV. In addition, ATIC~\cite{ATIC} and Fermi
LAT~\cite{Abdo:2009zk} have both reported anomalous structures in the
electron-plus-positron spectra up to around 800~GeV. These anomalies seem
to indicate a hitherto unknown source of primary electrons and positrons.
This has lead to
a lot of theoretical activity trying to explain both spectral features
within various dark matter models, while at the same time
avoiding severe constraints from current gamma-ray data, due to
radiation from the annihilation or decay products of the dark matter candidate.

In what follows, we will investigate how well the features of the
PAMELA and Fermi LAT data can be explained by decaying
gravitinos. Since there is some discrepancy between ATIC and Fermi LAT,
in particular at energies above 300 GeV, we
focus on the Fermi LAT data, due to the smaller errors. The
ATIC data, having a less smooth spectrum, cannot be fitted
as well as the Fermi LAT data in the present scenario.
In general, one can say that the ATIC data prefers a slightly harder
spectrum  than Fermi LAT, thus operators with more pronounced
electron flavour (see below) would be favored.

The main source of high-energy electrons and positrons in gravitino
decays with trilinear R-violating operators is the direct production
of charged leptons as occurs for $LL\bar E$ and $LQ\bar D$
operators. For a given gravitino mass and electron-flavour lepton
number violating operator we get a hard spectrum of direct electrons
and positrons, while for second and third generation lepton number
violating operators and the same gravitino mass we get softer spectra
from the decays of $\mu$ and $\tau$. In the case of hadronic tau
decays, as well as the direct production of quarks, we get many
lower-energy electrons from the decay of charged pions.

\begin{figure}[t]
  \includegraphics[width=17cm]{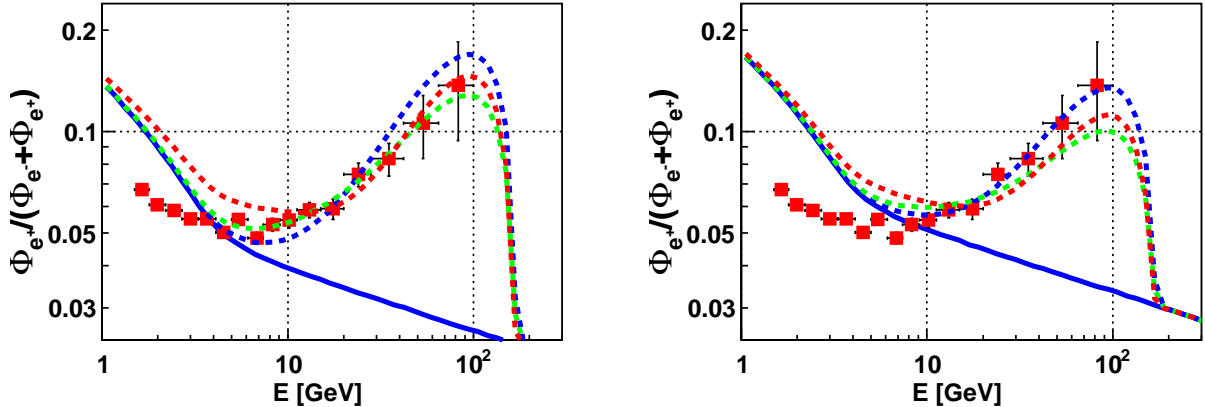}
  \caption{\it Data on positron fraction from PAMELA (red, with error bars)
compared to GALPROP background (solid blue) and fitted contributions from
$\mg=320~\text{GeV}$ gravitino decaying through an $L_1L_2\bar{E}_1$ operator
(dashed blue), an $L_1L_3\bar{E}_3$ operator (green) and an $L_1Q_2\bar{D}_2$
operator (red). The left panel shows the standard GALPROP background, in
the right panel the primary electron spectrum has been rescaled by a
factor 0.75.
}
\label{Fig:ep-320}
\end{figure}

To fit the PAMELA data with $LL\bar{E}$ operators is rather
straightforward; it is in fact possible to get a decent fit with any such operator,
provided the gravitino mass is sufficiently large (most require a
gravitino mass of at least 320~GeV). Also $LQ\bar{D}$ operators can be used
given a $L_1$ operator and a suitable gravitino mass. In Fig.~\ref{Fig:ep-320} we show
three examples where $L_1L_2\bar{E}_1$, $L_1L_3\bar{E}_3$ and
$L_1Q_2\bar{D}_2$ have been used to fit the PAMELA data by varying the
coupling for a $\mg=320~\text{GeV}$ gravitino, while assuming a common
mass of $m_{\rm SUSY}=1$~TeV for the other sparticles. The details of
the fits are given in Table~\ref{table:fit_PAMELA}.

\begin{table}[h]
\begin{center}
{\small
\begin{tabular}{|c|c|c|c|c|c|c|}
\hline
& \multicolumn{3}{|c|}{GALPROP background} &\multicolumn{3}{|c|}{Rescaled background}\\\hline
Coupling & $\lambda$ & $\tau~[10^{26} \text{ s}]$ &  $\chi^2_\text{PAM}/{\rm ndf}$ & $\lambda$ & $\tau~[10^{26} \text{ s}]$ & $\chi^2_\text{PAM}/{\rm ndf}$ \\\hline
$\lambda_{121}$ & $6.2\times 10^{-8}$ & 3.9 & 2.7  & $4.1\times 10^{-8}$ & 8.9 & 0.3\\
$\lambda_{133}$ & $8.1\times 10^{-8}$ & 2.3 & 0.3  & $5.5\times 10^{-8}$ & 5.0 & 1.4\\
$\lambda^\prime_{122}$ & $9.3\times 10^{-8}$ & 1.7 & 0.6  & $6.2\times 10^{-8}$ & 3.8  & 2.1\\
\hline
\end{tabular}
}
\caption{\label{table:fit_PAMELA} \it Best fit values for R-violating couplings
to the PAMELA positron fraction for $\mg=320$~GeV and
$m_\text{SUSY}=1$~TeV. The respective gravitino
lifetimes and the fit quality are also shown.
}
\end{center}
\end{table}

The background spectra of primary electrons from SNRs, as well as the
secondary electrons and positrons, have been calculated by {\sc
  GALPROP}. Which operator gives the best fit to data is sensitive to
the assumptions entering into the calculation.  If the standard {\sc
  GALPROP} parameters given above are used, as shown in
Fig.~\ref{Fig:ep-320} (left panel), operators with less direct
production of positrons and thus flatter spectra, $L_1L_3\bar{E}_3$
and $L_1Q_2\bar{D}_2$, give a better fit. On the other hand, a small
rescaling of the primary electrons by a factor 0.75 favors the steeper
rise given by the $L_1L_2\bar{E}_1$ operator (right panel).

Given this sensitivity to small changes in the background assumptions
for the PAMELA positron fraction, we will not discuss in further detail
which operators are more favored by PAMELA alone. Nor will we go to the other extreme and attempt
a global fit of all data, as this would require a sophisticated
treatment of experimental errors and the propagation model that go
beyond the scope of this Letter\footnote{For a detailed study of these
  issues, see~\cite{Barger:2009yt}.}. Instead, we adopt a simple
approach where we investigate whether it is possible to fit the Fermi
LAT data, and at the same time retain a reasonable fit quality to the
PAMELA data, allowing for small adjustments in the background assumptions.
In order to
simultaneously explain both data sets, we find that it is sufficient
to scale down the background of primary electrons somewhat compared to the
conventional GALPROP model. In what follows, we have adopted a simple
rescaling factor of 0.75.

Let us begin with some generic lessons learned concerning operator flavours
and gravitino masses, when attempting to fit both data-sets.  In order to fit
the Fermi LAT data alone, a minimal requirement is a rather high
gravitino mass, of at least $\mg\gtrsim 1.5$~TeV. However, for most
operators, even higher gravitino masses are required. Given such a
large gravitino mass, it is important not to have a too steep spectrum
in order to simultaneously explain the PAMELA rise, which takes place
at a much lower energy than the Fermi LAT excess. This can be best
achieved with a significant amount of tau flavour in the operator. We
find that operators of the type $L_iL_j\bar{E}_3$ are most
suitable for a simultaneous fit, since they require at least one
tau in the final state of the gravitino decay. At the same time,
operators of the type $L_iL_j\bar E _1$ seem excluded as an
explanation of both anomalies, due to their too steep spectra: the
magnitude of the coupling that is required in order to explain PAMELA
gives too large a contribution to the electron spectra at higher
energies. Finally, to have a sufficiently large contribution to the
high energy electron spectra, one either needs some electron flavour
in the operator, $i=1$ or $j=1$, or an even higher gravitino mass.

The qualitative description above is quantified in
Table~\ref{table:fit_Fermi}, which lists the operators we find capable of
simultaneously fitting both the Fermi LAT and the PAMELA data, {\it i.e.}\ in statistical terminology
operators with $\chi^2$ such that we fail to reject them at the 5\% significance level.
Table~\ref{table:fit_Fermi} also gives the $1\sigma$ error on the
gravitino mass; both the gravitino mass and the error are given from a two-parameter ($\lambda$ and $\mg$)
maximum likelihood fit to the Fermi LAT data.

Under the assumption
of a single coupling dominance, the value of the coupling
$\lambda$ does not affect the shape of the particle spectrum, but only
its normalization. As a result, the statistical error on the coupling for a given gravitino mass
is quite small, around 2--3\%, far smaller than the possible
systematic errors on the data normalization from the
energy scale in the experiments~\cite{Abdo:2009zk}. Therefore we only
give the value of the coupling for the best fit gravitino mass. The
behaviour of the coupling as a function of the gravitino mass around
the best fit, $\lambda\propto\mg^{-3.5}$, is clear from the gravitino
width dependence in the three-body decay, Eq.~(\ref{dec1}). We also
give the gravitino lifetime for the best fit gravitino mass and the
$\chi^2$ of the fit decomposed into the individual contributions from
the two data sets, divided by the degrees of freedom: for PAMELA $n=7$,
and for Fermi LAT $n-1=25$. Note that for the coupling $\lambda^\prime_{133}$
we have not conducted any fit of the gravitino mass. This is because it cannot simultaneously fit
both the Fermi LAT and the PAMELA data. It is included in Table~\ref{table:fit_Fermi}
merely for future reference.

\renewcommand\arraystretch{1.3}

\begin{table}[h]
\begin{center}
{
\begin{tabular}{|c|c|c|c|c|c|}
\hline
Coupling  &  $\mg\ [\text{TeV}]$ &  $\lambda$ at best fit  &  $\tau\ [10^{26} \text{ s}]$
&  $\chi^2_\text{PAM}$  &  $\chi^2_\text{Fermi}$  \\\hline
$\lambda_{123}$ & 1.8$^{+0.1}_{-0.2}$  &  $7.3\times 10^{-9}$ & 2.0 & 1.0 & 0.9 \\
$\lambda_{132}$ & 1.8$^{+0.1}_{-0.1}$  &  $6.9\times 10^{-9}$ & 2.3 & 1.7 & 1.1 \\
$\lambda_{133}$ & 1.8$^{+0.1}_{-0.3}$  &  $8.0\times 10^{-9}$ & 1.7 & 0.6 & 0.8 \\
$\lambda_{232}$ & 2.8$^{+0.4}_{-0.2}$  &  $1.7\times 10^{-9}$ & 1.5 & 1.6 & 1.1 \\
$\lambda_{233}$ & 3.6$^{+0.6}_{-0.3}$  &  $8.7\times 10^{-10}$ & 0.9 & 0.4 & 0.6 \\\hline
$\lambda^\prime_{133}$ & 1.8  &  $7.8\times 10^{-10}$ & 1.3 & 27 & 0.8 \\
\hline
\end{tabular}
}

\caption{\label{table:fit_Fermi} \it Operators for which the
  Fermi LAT and PAMELA electron and positron data can be fitted
  simultaneously. $\lambda$
  and $\tau$ are given for the central gravitino mass value for a given
  operator (see text). All other sparticle masses
  entering the calculation are set to 6 TeV. In the case of $\lambda^\prime_{133}$ the mass
  has not been fitted and the sparticle masses used here are 2 TeV (hence the smaller coupling).}
\end{center}
\end{table}

As seen in Table~\ref{table:fit_Fermi}, the best fits are given by $L_1L_3\bar E_3$ with a $1.8$ TeV gravitino and $L_2L_3\bar E_3$ with a $3.6$ TeV gravitino. These two fits are shown in Fig.~\ref{Fig:1800,LLE133}, where we also include a comparison to HESS data on the electron spectrum at even higher energies~\cite{Collaboration:2008aaa}.  It is, however, important to keep in mind that, even if an operator cannot fit the data by itself, it might still do so in combination with some other suitable operator.

\begin{figure}[t]
  \includegraphics[width=17cm]{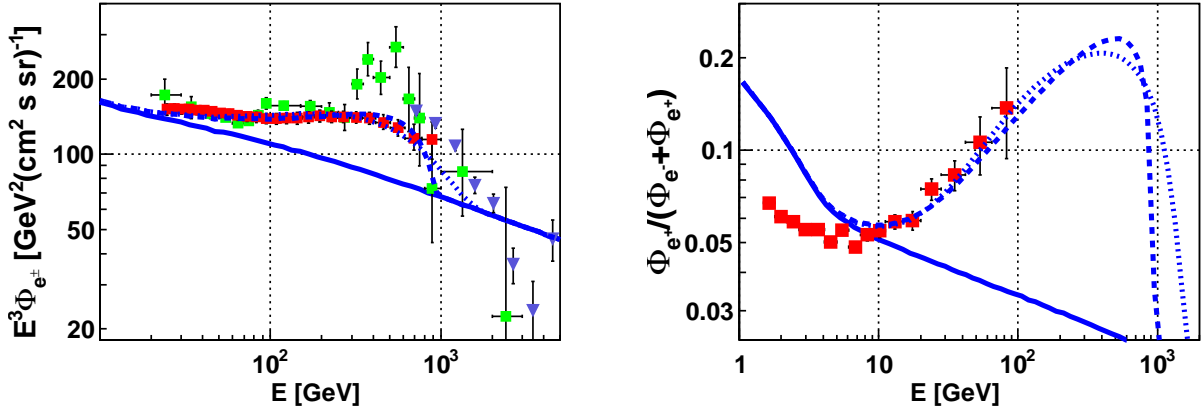}
  \caption{\it Left panel: fit of $L_1L_3\bar E_3$ operator with $\mg = 1.8$~TeV and $m_\text{SUSY}=6$~TeV (dashed blue), $L_2L_3\bar E_3$ operator with $\mg = 3.6$~TeV and $m_\text{SUSY}=6$~TeV (dotted blue) and {\sc GALPROP} background (solid blue) to electron-plus-positron spectrum from Fermi LAT (red, with error bars). Also shown is ATIC data (green, with error bars) and HESS data (violet inverted triangles, with error bars).  Right panel: data on positron fraction from PAMELA (red, with error bars) shown with {\sc GALPROP} background (solid blue) and the result of the fit for the $L_1L_3\bar E_3$ (dashed blue) and $L_2L_3\bar E_3$ (dotted blue) operators.
  }\label{Fig:1800,LLE133}
\end{figure}

If we were to consider the excess seen in the HESS spectrum
at around 1~TeV~\cite{Collaboration:2008aaa} in the fits, there would be some
preference for the higher ends of the gravitino mass regions;
for $L_1L_3\bar E_3$ and $L_1L_2\bar E_3$ one would need $\mg>2$ TeV,
while $L_2L_3\bar E_3$, due to its softer cut-off at high energy,
fits the HESS data more naturally.

So far we have exclusively fitted to the Fermi LAT data and ignored
the ATIC data.  It is of course possible to fit to the ATIC data
instead. As has been mentioned above, this favors a larger amount of
electron flavour in the operator and it turns out that the best
operator for the ATIC data is $L_2L_3\bar E_1$.
Since the ATIC data is less
smooth than the Fermi LAT data, it is not possible to fit it equally
well. The large error bars in the peak of the ATIC data also reduce
the tension between the electron-plus-positron data and the positron fraction data,
suggesting that fitting to Fermi LAT is the more reasonable and
constraining approach.

It is also possible to achieve a reasonable fit to the Fermi LAT data
with $LQ\bar D$ operators that contain $L_1$.  The best such
example is $L_1Q_3\bar D_3$, which is shown in
Fig.~\ref{Fig:1800,LQD133}; however, $LQ\bar D$
operators cannot fit simultaneously the PAMELA and Fermi LAT data,
due to positron excesses at low energies. We shall see below that
the interpretation in terms of $LQ\bar D$ operators is also
clearly excluded by the PAMELA antiproton data.

\begin{figure}[t]
  \includegraphics[width=17cm]{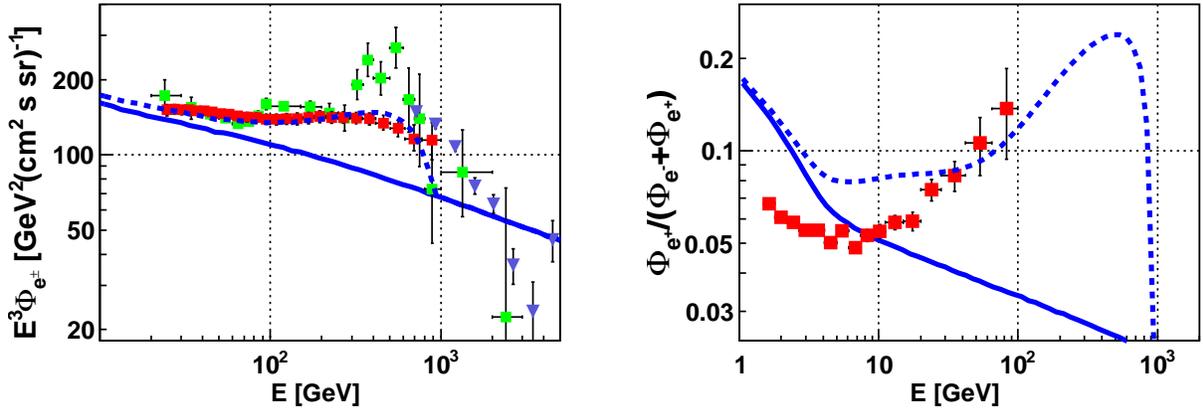}
  \caption{\it Similar to Fig.~\ref{Fig:1800,LLE133} for an $L_1Q_3\bar
  D_3$ operator with $\mg = 1.8$~TeV and $m_\text{SUSY}=2$~TeV.}
\label{Fig:1800,LQD133}
\end{figure}

It is interesting to note some important differences between this
scheme and dark matter annihilation scenarios: direct annihilation to
lepton pairs gives a large contribution of monochromatic electrons and
positrons in the CM frame. Even when galactic propagation is taken
into account, it is difficult to achieve the smooth excess over the
background spectrum that is observed by Fermi LAT. Consequently, dark
matter annihilation to an electron-positron pair seems largely
excluded as an explanation of the Fermi LAT~\cite{DMAnni} excess. For
gravitino decays, on the other hand, the three-body phase space can
naturally give rise to the smooth shape observed.

Let us finally comment on the consequences of the above for the
LHC. As can be seen in Table~\ref{table:fit_Fermi}, due to the large
gravitino masses and the $\mg^7$ dependence of the gravitino decay
width, the R-parity violating couplings need to be very small in order
to fit the data, $\mathcal{O}(10^{-9})$. This implies that, in such a
scheme, R-parity violation cannot be observed at the LHC unless the
NLSP is charged and can be stopped. The large sparticle masses
required also imply that the production cross section is very low, if
at all non-zero.

\subsection{Antiprotons}

In the case of $LQ\bar D$ and $\bar U \bar D\bar D$ operators, one
would also expect production of protons and antiprotons within the
quark jets. The PAMELA antiproton data \cite{PAMELApbar} give rather
tight constraints on all $LQ\bar D$ and $\bar{U}\bar{D}\bar{D}$
couplings for high gravitino masses. In Fig.~\ref{Fig:pbar,LQD} we
show the antiproton fraction data and compare to fits to the PAMELA
positron fraction (left) and the Fermi LAT electron-plus-positron flux
data (right). In both cases the $LQ\bar D$ operators that fit the
lepton data overproduce antiprotons, and are effectively ruled
out. From Fig.~\ref{Fig:pbar,LQD} we can also see that the pure
background is fairly consistent with the data, considering the potential for
large systematic errors. $LL\bar E$ operators on the other hand, do not produce
any antiprotons and are therefore not affected by these bounds.

\begin{figure}[t]
  \includegraphics[width=17cm]{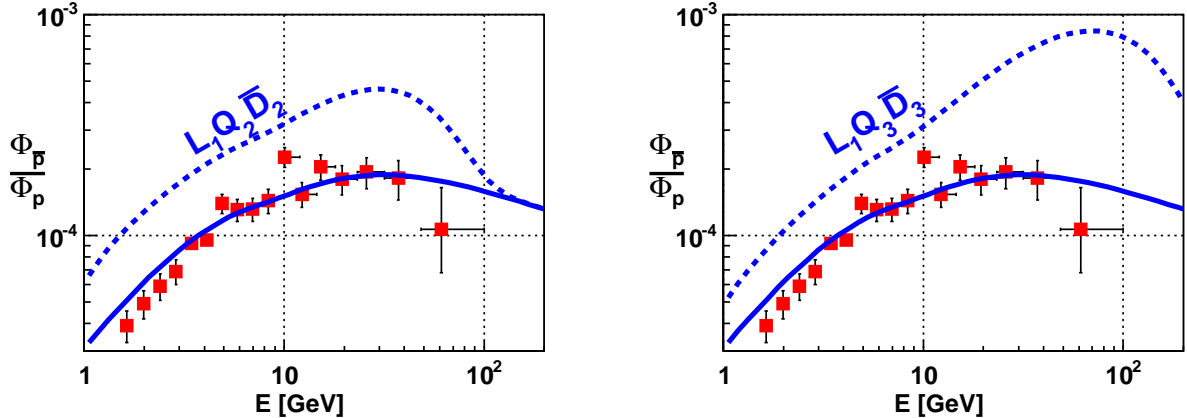}
  \caption{\it PAMELA data on the antiproton fraction (red, with error
  bars) compared to the {\sc GALPROP} background (solid blue) and the
  contribution from the best fit to the PAMELA positron fraction using
  the $L_1Q_2\bar D_2$ operator (left, dashed blue) and the best fit
  to the Fermi LAT electron-plus-positron spectrum with the
  $L_1Q_3\bar D_3$ operator (right, dashed blue). The parameters of
  the fits can be found in Tables~\ref{table:fit_PAMELA} (``rescaled background'') and
\ref{table:fit_Fermi}, respectively.
  }
  \label{Fig:pbar,LQD}
\end{figure}

\subsection{Maximal couplings}\label{Sec:MaxCoupling}

From the above discussion, it follows that the R-violating couplings
can be significantly constrained by the charged particle data. We determine such
limits by varying the gravitino mass and the coupling, while restricting the
operator to give a total flux that does not exceed any data point by more
than $3\sigma$.

To be conservative when setting the bounds there are no backgrounds
included in these calculations, except in the case of the positron
fraction where the electron background has to be included in order to
get a meaningful result; the positron background, however, is not
included.  In contrast to the fits, we are using here the full spectra
of the PAMELA data.  Given the relatively small experimental
uncertainties at lower energies, and having excluded the background
from the calculation, we should still get robust constraints, despite
the solar modulation uncertainties at the lowest energies.

\begin{figure}[t]
  \includegraphics[width=17cm]{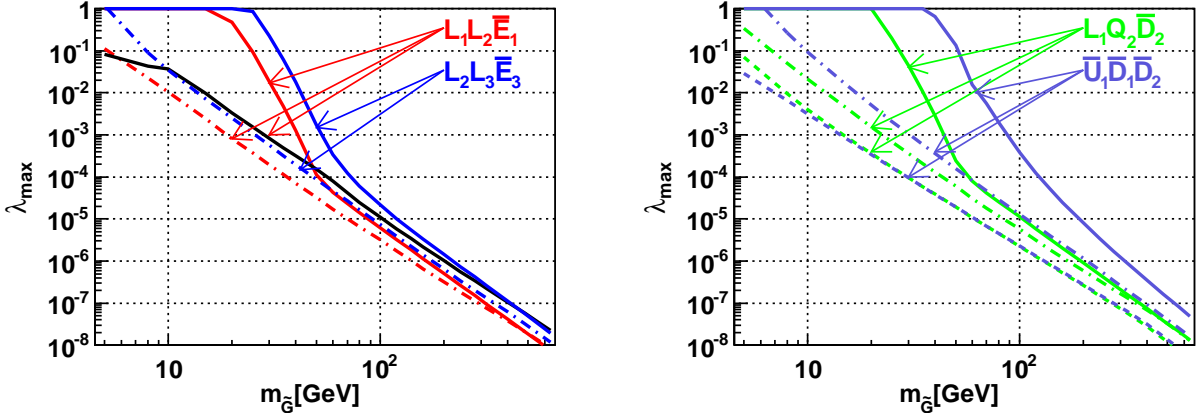}
  \caption{\it Maximum allowed couplings $\lambda_{\max}$ versus gravitino mass, derived
    from the Fermi LAT electron-plus-positron data (solid), the PAMELA
    positron fraction data (dot-dashed) and antiproton data
    (dashed). We show bounds for the $L_1L_2\bar E_1$ (red),
    $L_2L_3\bar E_3$ (blue), $L_1Q_2\bar D_2$ (green) and $\bar
    U_1\bar D_1\bar D_2$ (violet) operators. For comparison, the
    constraint derived from the EGRET gamma ray data in~\cite{BLOR} is shown for
    $L_1L_2\bar E_1$ (solid black). }
\label{Fig:lambda}
\end{figure}

We present the resulting bounds for a representative selection of operators in Fig.~\ref{Fig:lambda}. For comparison we also show the limits on the coupling from the EGRET gamma-ray data for $L_1L_2\bar E_1$. For details on the derivation of this limit and comparisons between gamma-ray constraints on different operators, see~\cite{BLOR}.  In Fig.~\ref{Fig:lambda} we use $m_{SUSY} = 1$ TeV, for other sparticle masses one can use the scaling $\lambda_{max}\propto m_{SUSY}^2$. The range in gravitino mass shown is chosen to be comparable with earlier work~\cite{BLOR}, and to cover the range where RPV decays at the LHC might be possible.

For $LL\bar E$ operators, the positron fraction data gives the most constraining bounds for medium to high gravitino masses. As expected, this bound is stronger for operators with direct positron production, {\it e.g.} $L_1L_2\bar E_1$, as compared to indirect production, via $L_2L_3\bar E_3$. At low gravitino masses, the operators that allow radiative loop decays, such as $L_1L_2\bar E_1$, will be mainly
constrained by the gamma-ray data, while at very high masses the constraints from the electron-plus-positron data is competitive. For $LQ\bar D$ operators we observe a similar behaviour, however, in this case the antiproton data provide even stronger bounds. Finally, $\bar U\bar D\bar D$ operators have no radiative decays and no direct electron/positron production, but produce both positrons and gammas from pion decays, resulting in bounds similar to, but weaker than, the bounds on the $LQ\bar D$ operators. As expected the antiproton data is more restrictive; the constraint from antiprotons turns out to be almost identical to the one for $LQ\bar D$ operators. This is something of a numerical accident, $\bar U\bar D\bar D$ operators give more antiprotons, but the $LQ\bar D$ operators have roughly twice the gravitino decay width for the same coupling strength, and these effects cancel each other out.

Overall, we see that for gravitino masses smaller than the values
required to fit PAMELA and Fermi LAT, the allowed couplings can be
significantly larger and will result in observable R-violation at the LHC.

\section{Photon Spectra}
\label{sec:photons}
\setcounter{equation}{0}

In this section we discuss the photon spectra from gravitino decays,
and the possibility of either excluding or strengthening the gravitino
interpretation of the electron and positron excesses using current and
future gamma-ray measurements.

The data we use for comparison is the extragalactic spectrum from EGRET
as calculated by~\cite{reEGRET}. The choice of this data over the
first EGRET analysis~\cite{EGRET} is due to the improved background modeling
in~\cite{reEGRET}.

\subsection{Model}

There are three main contributions to the gamma-ray spectrum from
gravitino decays: there is direct photon production in the radiative
loop decays (giving a monochromatic component), there are photons from
internal bremsstrahlung of charged particles, and there are photons
from pion decays. In this case as well, we let {\sc
  PYTHIA}~\cite{PYTHIA} simulate the gravitino decay including all
relevant decay channels. The photon spectrum reaching the earth can be
further divided into two components; one extragalactic part that has
been red-shifted from its energy in the gravitino rest frame and one
part from the galactic halo.

The extragalactic contribution to the total flux is given by \cite{IbTran}
\begin{equation}
\left[E^2\frac{dJ}{dE}\right]_\text{EG}
=\frac{2E^2}{\mm}C_\gamma\int^\infty_1 dy \frac{dN_\gamma}{d(Ey)}
\frac{y^{-3/2}}{\sqrt{1+\kappa y^{-3}}},
\label{Eq:EG,flux}
\end{equation}
where $y= 1+z$, $z$ being the redshift, and $dN_\gamma/dE$ the
gamma ray spectrum from the gravitino decay in its rest frame, as
calculated by PYTHIA. The constants $C_\gamma$ and $\kappa$ are given
by
\begin{equation}
C_\gamma=\frac{\Omega_{\tilde G}\rho_c}
{8\pi\tau_{\tilde G} H_0 \Omega_M^{1/2}}
\quad {\rm and} \quad \kappa=\frac{\Omega_\Lambda}{\Omega_M}.
\end{equation}

The halo component has been calculated using the NFW \cite{NFW} dark
matter halo profile of Eq.~(\ref{Eq:NFW}), which has been averaged
over all directions such that $|b|\geq 10^{\circ}$, where $b$ denotes
galactic latitude. This exclusion is done in order to avoid the difficult
background of strong gamma ray sources within the galactic disc, and
mimics what is done in the analysis of the EGRET data~\cite{reEGRET}
that we will be comparing to\footnote{There are slight differences in
the exclusion bands used in the various data sets from gamma-ray
experiments. Since we are not looking for point sources, and we are
using most of the sky, this should have little effect on the
final result.}.  The halo component of the flux is then given as
\begin{equation}
\left[E^2\frac{dJ}{dE}\right]_\text{Halo}
=\frac{E^2}{\mm}\frac{dN_\gamma}{dE}
\frac{1}{4\pi\tg}\int_\text{los}\rho_\text{Halo}{\vec{l}}d\vec l ,
\label{Eq:Halo,flux}
\end{equation}
where $\int_{los}\rho_{Halo}{\vec{l}}d\vec l $ is the
line-of-sight integral averaged over the specified part of the halo profile.
The resulting gamma ray
spectra have also been smoothed to account for the effects of a
detector with energy resolution of 15\%, again on the basis of the
properties of the EGRET analysis.

\begin{figure}[!h]
 \includegraphics[width=17cm]{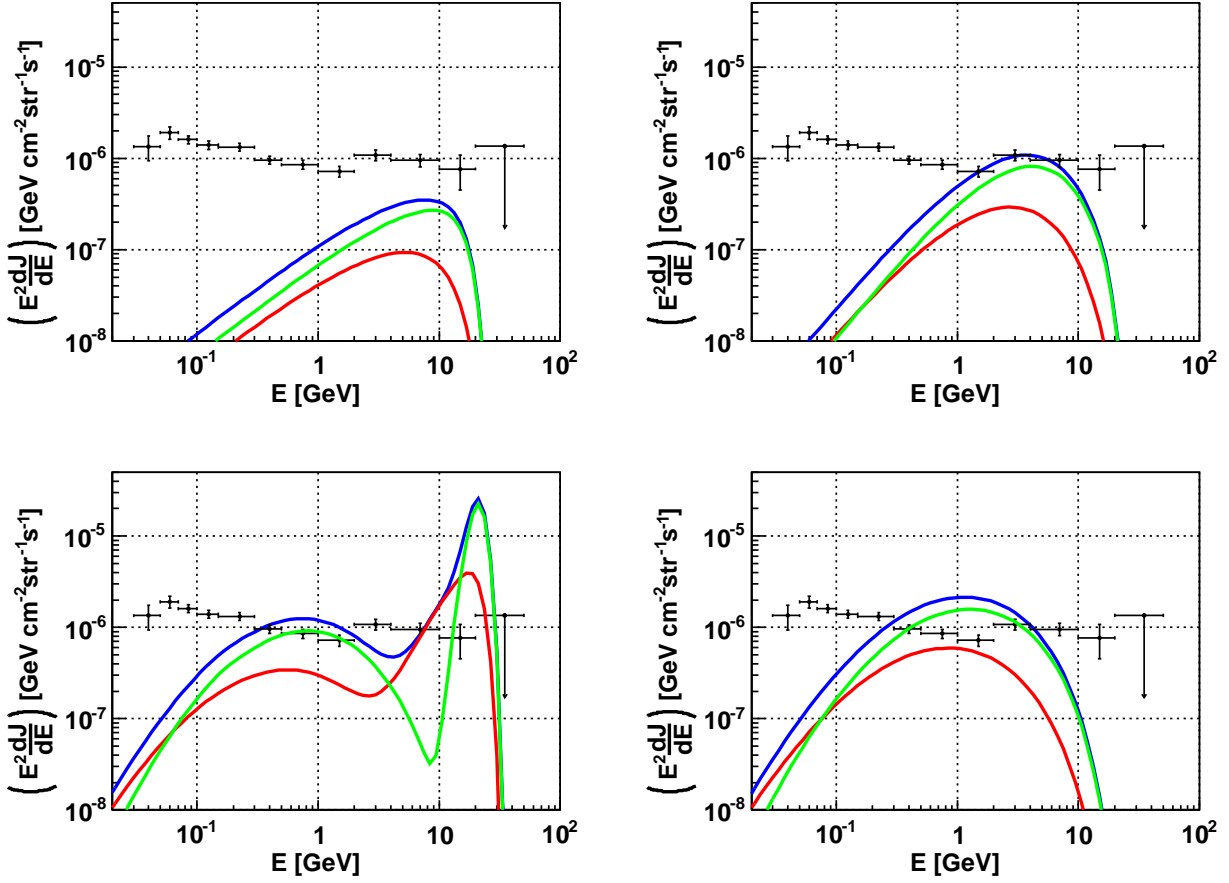}
 \caption{\it Photon spectra for four representative R-violating
 operators: $L_1L_2\bar E_1$ (top left), $L_1L_2\bar E_3$ (top right),
 $L_1Q_3\bar D_3$ (bottom left) and $\bar U_2\bar D_1\bar D_2$ (bottom
 right), for the following set of parameters: $\mg = 40$~GeV,
 $M_\text{SUSY}=200$~GeV, and $\lambda=10^{-5}$.  We show the
 extragalactic (red) and halo (green) contributions, as well as the
 total (blue). Also shown is the EGRET data.
}
\label{fig:photons}
\end{figure}

In  Fig.~\ref{fig:photons} we show photon spectra for
different flavour combinations in the R-violating
couplings, demonstrating the variety of possible gamma-ray spectral
shapes that can be generated.
To illustrate the potential observable
consequences of decaying gravitino dark matter in a gamma-ray
experiment, we pick the R-parity violating couplings in such a way
that the spectrum is close to the EGRET measurement, and for comparison
we also show the EGRET data~\cite{reEGRET}.

From Fig.~\ref{fig:photons} we observe that the two-body decays (when
present, such as in the lower left plot) tend to dominate and the result is a monochromatic line at
$\mg/2$. In an experiment this would then show up
as a broad peak; however, a gravitino mass could
potentially be identified if sufficient statistics were available, the accuracy limited
mostly by the experimental resolution. Furthermore, we observe the following:
\begin{itemize}
\item[(i)] When all decay products are electrons and muons (top-left
  panel) practically all gamma radiation comes from internal
  bremsstrahlung. This radiation is quite hard and has a sharp cut-off
  at $\mg/2$.
\item[(ii)] For $LL\bar{E}$ operators with some $\tau$ flavour (top-right panel) gamma
  rays come from both internal brems\-strahlung and decaying mesons from
  tau decays, mostly $\pi^0$, giving a softer spectrum as compared to
  only bremsstrahlung, for the same gravitino mass.
\item[(iii)] In the case of $L_iQ_3\bar{D}_3$ operators (bottom-left
  panel) we get a combination of photons from loop and
  tree-level decays (the latter give photons through decaying mesons). The
  heavier the particle in the loop, the larger the loop contribution;
  $L_iQ_3\bar{D}_3$ has the most pronounced loop contribution of all
  operators, due to the high mass of the b-quark \cite{LOR}.
\item[(iv)] Finally, for $\bar{U}\bar{D}\bar{D}$ couplings
  (bottom-right panel), the only sizable contribution arises through
  meson decays, resulting in a soft spectrum.
\end{itemize}

\subsection{Predictions from PAMELA and Fermi LAT}
\label{sec:PAMELA-ATIC}
\setcounter{equation}{0}

We can now make predictions for what should be observed in gamma-ray
experiments assuming that the PAMELA and Fermi LAT data discussed in
the previous section are explained by trilinear R-parity violating
operators. As we have seen, the only option is to use $LL\bar E$
operators and large gravitino masses, above 1~TeV. In this range,
three-body decays are likely to completely dominate, implying the
absence of an observable peak of monochromatic photons. However, with such high
gravitino masses, a considerable photon flux is still expected from
bremsstrahlung and pion decays.

In Fig.~\ref{Fig:gamma,LLE-133,233} we show the predictions from two of the fits in Section~\ref{sec:charged} ($L_1L_3\bar E_3$ and $L_2L_3\bar E_3$) compared to the EGRET data.  It is clear that both predictions are compatible with the data because the large gravitino mass implies that the dark matter spectrum lies outside the EGRET sensitivity range.  However, new Fermi LAT data on gamma rays is expected to cover at least parts of this region, and our best fit models predict a broad spectral feature whose position is correlated to the gravitino mass. Assuming a power-law like continuation of the EGRET data this feature may be observed, or our present models excluded. If both the rise and the cut-off of the feature is observed, an estimate of a probable gravitino mass may be made, and the type and size of the R-violating coupling could be further constrained.

\begin{figure}[t]
  \includegraphics[width=17cm]{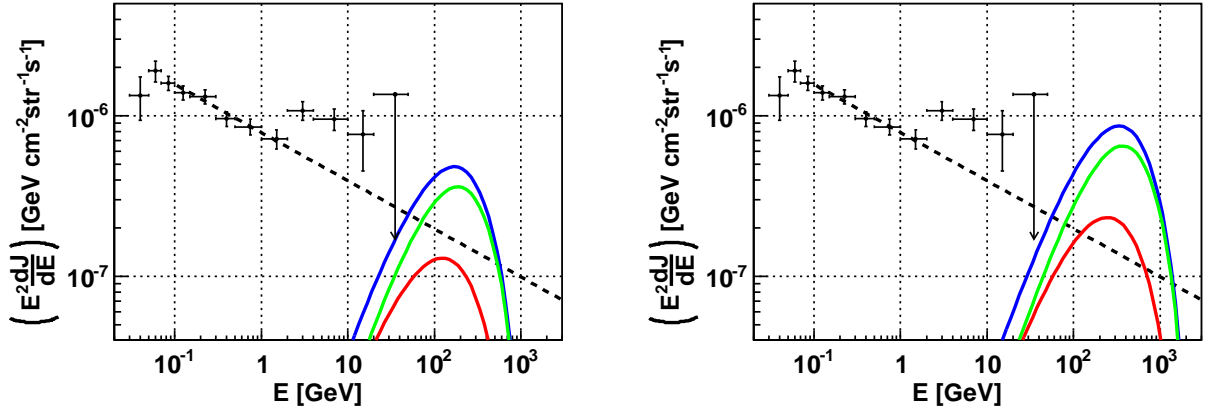}
  \caption{\it Photon spectra for the best-fit models of
    Fig.~\ref{Fig:1800,LLE133}; $\lambda_{133}=8.0\times 10^{-9}$, $\mg = 1.8$ TeV (left) and $\lambda_{233}=8.7\times 10^{-10}$, $\mg = 3.6$ TeV (right). We show the extragalactic (red) and halo (green) contributions,
    as well as the total (blue). For comparison we also
    give the EGRET data,
    and a power-law continuation of the data in the interval [0.05,2]~GeV (dashed line).}
   \label{Fig:gamma,LLE-133,233}
\end{figure}

\section{Neutrinos}
\label{sec:neutrinos}
\setcounter{equation}{0}

In addition to the photon flux, the two-body decay mode also generates a monochromatic neutrino flux in the gravitino rest frame.  Moreover, in the three-body decays, neutrinos will be produced both in the initial decay, and in subsequent decays of unstable leptons and hadrons.  Neutrinos, like photons, travel essentially undeflected from the point of production. However, they undergo oscillations, changing from one flavour to another over distances which are small compared to galactic scales.

Ignoring possible CP violation in the neutrino sector, the transition probabilities can be expressed in terms of the mixing matrix $U$ as
\begin{equation}
P(\nu_\alpha\to\nu_\beta) =\sum_{k=1}^3(U_{\alpha k}U_{\beta k})^2.
\end{equation}
For the mixing we adopt the parameters of \cite{Strumia:2006db}, corresponding
to maximal mixing between the $\mu$ and $\tau$ flavours,
$\sin^2\theta_{23}=0.5$, and for the remaining mixings
$\sin^2\theta_{12}=0.304$ and $\sin^2\theta_{13}=0.01$.  This gives:
\begin{equation} \begin{split}
    &P(\nu_e\leftrightarrow\nu_e) =0.56\,, \\
    &P(\nu_e\leftrightarrow\nu_{\mu})=P(\nu_e\leftrightarrow\nu_{\tau}) =0.22\,, \\
    &P(\nu_{\mu}\leftrightarrow\nu_{\mu})=P(\nu_{\mu}\leftrightarrow\nu_{\tau}) =P(\nu_{\tau}\leftrightarrow\nu_{\tau}) =0.39\,.  \end{split} \label{Eq:oscprob}
\end{equation}
The effect of the above is to erase flavour specific information from the decay in the neutrinos that reach us. However, it also guarantees the presence of muon neutrinos in the flux, which, for the near future, seems to be our best hope for detection through the production of muons in the scattering off nucleons in experiments such as IceCube.  For electron neutrinos, the resulting electron energy in the present model is too low for reconstruction of the neutrino direction, making a rejection of the enormous background from cosmic ray showers difficult, while tau neutrinos contribute to the muon flux with a fraction corresponding to the tau branching ratio into muons.

Due to the similarities in the propagation, the neutrino flux is calculated in analogy with the gamma ray flux, {\it i.e.}, by the use of Eqs.~(\ref{Eq:EG,flux}) and (\ref{Eq:Halo,flux})\footnote{For detailed discussions of neutrinos from annihilating and decaying dark matter, see~\cite{Yuksel:2007ac} and~\cite{PalomaresRuiz:2007ry,Covi:2009xn}, respectively.}, taking into account neutrino oscillations. The resulting neutrino spectra have then been smoothed to mimic a 10\% energy resolution on muons in a detector. This smoothening causes the energy spectra to exceed the theoretical upper limit of $\mg/2$. In this simple model we have ignored the energy lost to the accompanying hadronic shower, the inelasticity of the event, which should further smear the spectrum toward lower energies. We also restrict ourselves to through-going muons, and ignore the muon energy loss in matter. For a more thorough treatment of these effects for generic models of decaying dark matter, see~\cite{Covi:2009xn}.

\begin{figure}[ht]
\begin{center}
  \includegraphics[width=0.45\textwidth]{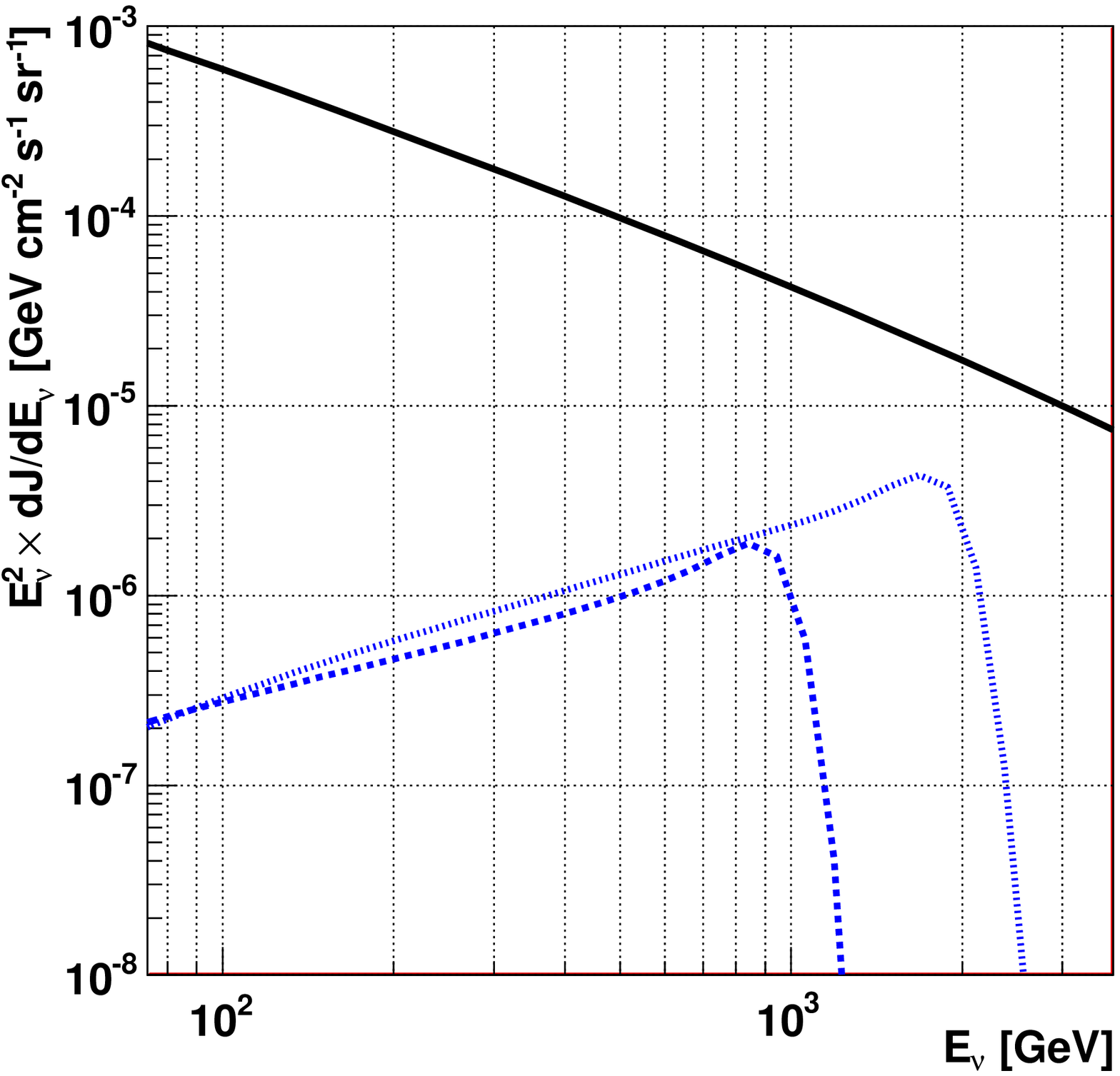}
  \includegraphics[width=0.45\textwidth]{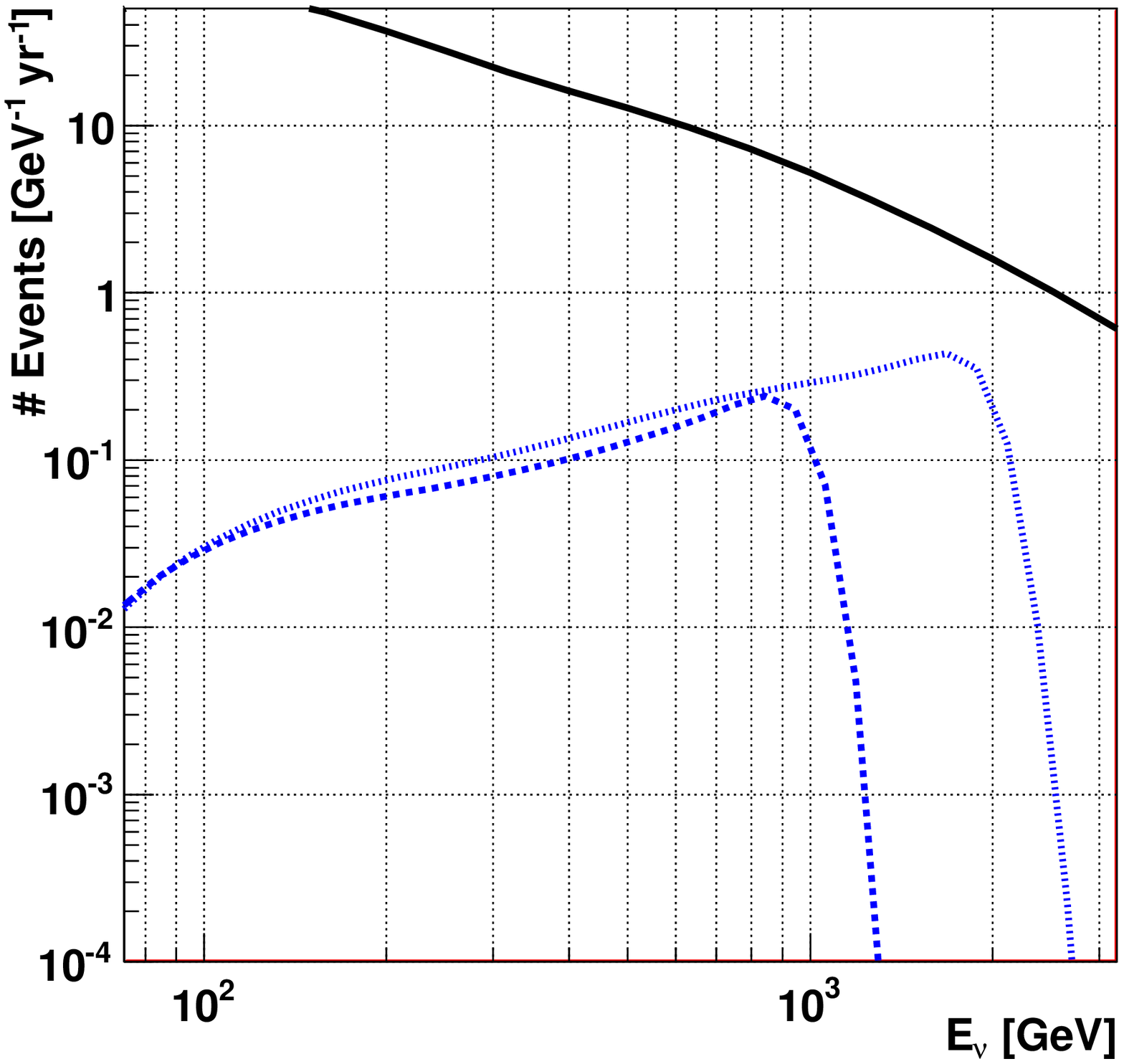}
  \caption{{\it Neutrino $(\nu_\mu+\bar{\nu}_\mu)$ spectrum (left) and expected number of events in IceCube (right) for best fit models (blue)  discussed in Section~\ref{sec:charged} (dashed: $\lambda_{133}=8.0\times 10^{-9}$, $\mg = 1.8$ TeV, dotted: $\lambda_{233}=8.7\times 10^{-10}$, $\mg = 3.6$ TeV), compared to atmospheric neutrino flux (black, solid) .
}}\label{Fig:nu}
\end{center}
\end{figure}

The resulting muon neutrino spectrum is shown in Fig.~\ref{Fig:nu} (left) for the best-fit models shown in Fig.~\ref{Fig:1800,LLE133}. We compare this signal to the expected atmospheric neutrino background spectrum taken from~\cite{Honda:2006qj}. Figure~\ref{Fig:nu} (right) also shows the expected number of signal and background events per year in the IceCube experiment, calculated by folding the neutrino spectra with the effective area of a completed IceCube experiment averaged over the northern hemisphere~\cite{Montaruli:2009kv}.

It is worth noting that the similarity at low neutrino energies and the difference in cut-off between the two models shown in Fig.~\ref{Fig:nu} is a direct consequence of the fit to the Fermi LAT electron-plus-positron data and its limited reach in energy shown in Fig.~\ref{Fig:1800,LLE133}. The model with larger gravitino mass fits the HESS data beyond Fermi LAT better. Thus, observing a cut-off in the Fermi LAT excess will predict the position of the cut-off in the neutrino spectrum in these models and {\it vice versa}.

We can see from Fig.~\ref{Fig:nu} (left) that neutrinos directly produced from the three-body $LL\bar{E}$ decays give an enticing peak in the neutrino flux around $\mg/2$. However, large atmospheric backgrounds make the detection of these neutrinos very challenging. For IceCube we expect ${\mathcal O}(500)$ upward going muons per year in total from these neutrinos for the best case scenario with the largest gravitino mass. This seems marginal when compared to an expected total atmospheric background of 25k events per year.  When using the spectral feature of the flux, the numbers are more encouraging, we find $S/\sqrt{B}=5.4$ and $S/B=0.19$ for events in the energy interval $1.6-2.0$~TeV.

One possible route to improvements is to look instead at downward going muons where efficient rejection of muon neutrinos produced in cosmic ray events may become feasible with the IceCube DeepCore detector~\cite{Schonert:2008is,Wiebusch:2009jf}. This was recently discussed for leptophilic dark matter in~\cite{Mandal:2009yk}.

\section{Summary-outlook}
\label{sec:conclusions}
\setcounter{equation}{0}

In this work, we have studied the expected charged particle, photon and neutrino spectra of slowly decaying gravitino dark matter within the framework of supersymmetric models with explicitly broken R-parity through trilinear operators. We identify couplings that generate spectra with a distinct behaviour that may reproduce recent experimental photon and (anti-)fermion data.  Among others, we find the following:
\begin{itemize}
\item
The cosmic ray electron and positron data, as reported by the PAMELA Collaboration, can be easily reproduced via $LL\bar E$ and $LQ\bar D$ operators. Couplings with significant electron flavour give rise to a hard spectrum, in contrast to the case where most electrons come from $\mu$ and $\tau$ decays.  In the case of tau as well as quark final states, low energy electrons are produced from the decay of charged pions. We find that most operators require a gravitino mass of at least 320~GeV to explain the PAMELA data.
\item
In order to fit the Fermi LAT data alone, we need even higher gravitino masses, of at least $\mg\gtrsim 1.5$~TeV (for most operators significantly larger).  In order to simultaneously account for the PAMELA rise, which takes place at a much lower energy, such gravitino masses also require a part of the spectrum that is not too hard, favoring operators of the $L_iL_j\bar{E}_3$ type. Sufficiently large contributions to the high energy electron spectrum can then be obtained either via some electron flavour in the operator ($i=1$ or $j=1$) or by going to even higher gravitino masses.  The decaying dark matter scenario discussed in this Letter has a significant advantage over dark matter annihilation scenarios, where it is hard to achieve a spectrum that is smooth enough to explain the electron excess observed by Fermi LAT.
\item
$L_1Q_j\bar D_k$ operators may give a reasonable fit to the Fermi LAT data alone, but cannot then fit the PAMELA data due to positron excesses at low energies and the  over-production of antiprotons.  In fact, the PAMELA antiproton data give very tight constraints on all $LQ\bar D$ and $\bar{U}\bar{D}\bar{D}$ couplings for high gravitino masses.
\item
For the range of gravitino masses that can match the PAMELA and Fermi LAT data, three-body decays are likely to dominate, implying the absence of the observable peak of monochromatic photons found in two-body radiative gravitino decays. Large photon fluxes are still expected, due to bremsstrahlung off charged particles and pion decays. Such fluxes are a necessary consequence of our scenario and should be detectable by Fermi LAT for parameters that can explain the charged particle excesses, meaning that Fermi LAT will be capable of supporting or
ruling out this explanation of the electron and positron anomalies.
\item
Neutrinos directly produced from the three-body decays have a quite
sharp peak around $\mg/2$, which, with the sharp rise in effective area with energy, gives some hope of future detection at IceCube. In particular, despite only measuring the muon energy from the neutrino interaction, this may lead to a better gravitino mass determination than what is possible from bremsstrahlung photons.
\end{itemize}

Summarising, it is interesting to observe that our predictions on the basis of the PAMELA and Fermi LAT data are very restrictive, and imply that in the future this scenario will either be confirmed, or very strict bounds will be imposed on R-violating couplings for heavy gravitino masses.  It is also interesting to note that R-violating couplings too small to detect at the LHC, can be probed through the study of photon and (anti-)particle spectra.

\vspace*{0.2 cm}

{\bf Acknowledgements.}
The work of NEB and PO has been supported by the Research Council of Norway.
SL has been funded by the FP6 Marie Curie Excellence Grant MEXT-CT-2004-014297
and also acknowledges support  by the European Research and
Training Network UniverseNet, MRTPN-CT-2006 035863-1. ARR is grateful for
financial support from the Swedish Research Council (VR) through the Oskar Klein
Centre and thanks Chad Finley for enlightening discussions.


\end{document}